# Duality between Packings and Coverings of the Hamming Space

**Gérard Cohen**

Département Informatique
Ecole Nationale Supérieure des Télécommunications
46 rue Barrault, 75634 Paris, FRANCE
*cohen@enst.fr*

**Alexander Vardy**

Department of Electrical and Computer Engineering
Department of Computer Science and Engineering
Department of Mathematics
University of California San Diego
9500 Gilman Drive, La Jolla, CA 92093, U.S.A.
*vardy@kilimanjaro.ucsd.edu*

November 4, 2018

*Dedicated to the memory of Jack van Lint*

### Abstract

We investigate the packing and covering densities of linear and nonlinear binary codes, and establish a number of duality relationships between the packing and covering problems. Specifically, we prove that if almost all codes (in the class of linear or nonlinear codes) are good packings, then only a vanishing fraction of codes are good coverings, and vice versa: if almost all codes are good coverings, then at most a vanishing fraction of codes are good packings. We also show that any *specific* maximal binary code is either a good packing or a good covering, in a certain well-defined sense.

Supported in part by the David and Lucile Packard Fellowship and by the National Science Foundation.

# 1. Introduction

Let $\mathbb{F}_2^n$ be the vector space of all the binary $n$-tuples, endowed with the Hamming metric. Specifically, the **Hamming distance** $d(x, y)$ between $x, y \in \mathbb{F}_2^n$ is defined as the number of positions where $x$ and $y$ differ. A **binary code** of length $n$ is a subset of $\mathbb{F}_2^n$, while a **binary linear code** of length $n$ and dimension $k$ is a $k$-dimensional subspace of $\mathbb{F}_2^n$. Since we are concerned *only* with binary codes in this paper, we henceforth omit the "binary" quantifier throughout. The **minimum distance** $d$ of a code $\mathbb{C} \subseteq \mathbb{F}_2^n$ is defined as the minimum Hamming distance between distinct elements of $\mathbb{C}$. The **covering radius** of $\mathbb{C}$ is the smallest integer $R$ such that for all $x \in \mathbb{F}_2^n$, there is a $y \in \mathbb{C}$ with $d(x, y) \leqslant R$. For all other notation from coding theory, we refer the reader to the book of van Lint [8]. Van Lint [8, p.34] calls the covering radius the "counterpart of minimum distance." Indeed, the trade-off between the parameters $|\mathbb{C}|$, $n$, $d$, and $R$ is one of the fundamental problems in coding theory.

Let $\mathscr{C}(n, M)$ denote the set of all codes $\mathbb{C} \subseteq \mathbb{F}_2^n$ with $|\mathbb{C}| = M$. Thus $|\mathscr{C}(n, M)| = \binom{2^n}{M}$. Similarly, let $\mathscr{L}(n, k)$ denote the set of all linear codes of length $n$ and dimension $k$. Thus the cardinality of $\mathscr{L}(n, k)$ is given by $|\mathscr{L}(n, k)| = \prod_{i=0}^{k-1} (2^n - 2^i)/(2^k - 2^i)$. We will be interested in questions of the following kind. Given a property **P** which determines the packing or covering density of a code, what fraction of codes in $\mathscr{C}(n, M)$ and/or $\mathscr{L}(n, k)$ have this property? Moreover, how does this fraction behave as $n \to \infty$? Our main results are curious duality relationships between such packing and covering problems. In particular, we show that:

- ✧ *Any maximal code is good. That is, any specific maximal code $\mathbb{C} \subseteq \mathbb{F}_2^n$ is either a good packing or a good covering, in a certain well-defined sense.*

- ❋ *If almost all codes in $\mathscr{C}(n, M)$ are good coverings, then almost all codes in $\mathscr{C}(n, M+1)$ are bad packings. Vice versa, if almost all codes in $\mathscr{C}(n, M+1)$ are good packings, then almost all codes in $\mathscr{C}(n, M)$ are bad coverings.*

- ❋ *The same is true for linear codes. That is, ❋ holds with $\mathscr{C}(n, M)$ and $\mathscr{C}(n, M+1)$ replaced by $\mathscr{L}(n, k)$ and $\mathscr{L}(n, k+1)$, respectively.*

The definition of what we mean by "good packing" and "good covering" is given in the next section. Precise statements of ✧ and ❋, ❋ may be found in §3 and §4, respectively.

# 2. Definitions

The **covering density** of a code $\mathbb{C} \subseteq \mathbb{F}_2^n$ is defined in [3] as the sum of the volumes of spheres of covering radius $R$ about the codewords of $\mathbb{C}$ divided by the volume of the space:

$$\mu(\mathbb{C}) \stackrel{\text{def}}{=} \frac{\sum_{c \in \mathbb{C}} |B_R(c)|}{|\mathbb{F}_2^n|} = \frac{|\mathbb{C}| V(n, R)}{2^n}$$

where $B_r(x) = \{y \in \mathbb{F}_2^n : d(x, y) \leqslant r\}$ is the sphere (ball) of radius $r$ centered at $x \in \mathbb{F}_2^n$ and $V(n, r) = \sum_{i=0}^{r} \binom{n}{i}$ is the volume (cardinality) of $B_r(x)$. We find it extremely convenient to extend this definition of density to arbitrary radii as follows.

**Definition 1.** *Given a code $\mathbb{C} \subseteq \mathbb{F}_2^n$ and a nonnegative integer $r \leqslant n$, the $r$-**density of** $\mathbb{C}$ is defined as*

$$\varphi_r(\mathbb{C}) \stackrel{\text{def}}{=} \frac{\sum_{c \in \mathbb{C}} |B_r(c)|}{|\mathbb{F}_2^n|} = \frac{|\mathbb{C}| \, V(n,r)}{2^n} \qquad (1)$$

Many well-known bounds on the packing and covering density of codes can be concisely stated in terms of the $r$-density. For example, if $R$, $d$, and $t = \lfloor (d-1)/2 \rfloor$ denote the covering radius, the minimum distance, and the packing radius, respectively, then

*Sphere-packing bound:* $\varphi_t(\mathbb{C}) \leqslant 1$ for all $\mathbb{C} \subseteq \mathbb{F}_2^n$ \qquad (2)

*Sphere-covering bound:* $\varphi_R(\mathbb{C}) \geqslant 1$ for all $\mathbb{C} \subseteq \mathbb{F}_2^n$ \qquad (3)

The classical Gilbert-Varshamov bound [8] asserts that for all $n$ and $d \leqslant n$, there exist codes in $\mathscr{C}(n, M)$ whose minimum distance $d$ satisfies $M \geqslant 2^n/V(n, d-1)$. Equivalently

*Gilbert-Varshamov bound:* $\forall n, \forall d \leqslant n$, there exist $\mathbb{C} \subseteq \mathbb{F}_2^n$, such that $\varphi_{d-1}(\mathbb{C}) \geqslant 1$ \qquad (4)

Recently, this bound was improved upon by Jiang and Vardy [7] who proved that for all sufficiently large $n$ and all[*] $d \leqslant 0.499n$, there exist codes $\mathbb{C} \subset \mathbb{F}_2^n$ with minimum distance $d$ such that $|\mathbb{C}| \geqslant cn \, 2^n/V(n, d-1)$, where $c$ is an absolute constant. Equivalently

$\exists c > 0, \forall n \geqslant n_0, \forall d \leqslant 0.499n$, there exist $\mathbb{C} \subseteq \mathbb{F}_2^n$, such that $\varphi_{d-1}(\mathbb{C}) \geqslant cn$ \qquad (5)

The best known existence bounds for covering codes can be also expressed in terms of the $r$-density, except that one should set $r = R$ rather than $r = d - 1$. Thus

$\forall n, \forall R < n/2$, there exist linear $\mathbb{C} \subseteq \mathbb{F}_2^n$, such that $\varphi_R(\mathbb{C}) \leqslant n^2$ \qquad (6)

$\forall n, \forall R < n/2$, there exist $\mathbb{C} \subseteq \mathbb{F}_2^n$, such that $\varphi_R(\mathbb{C}) \leqslant (\ln 2)n$ \qquad (7) where the first result is due to Cohen [4] while the second is due to Delsarte and Piret [5]. Motivated by (4)–(7), we introduce the following definition.

**Definition 2.** *Let $f(n)$ be a given function, and let $\mathbb{C} \subseteq \mathbb{F}_2^n$ be a code with minimum distance $d$ and covering radius $R$. We say that $\mathbb{C}$ is an $f(n)$-**good packing** if $\varphi_{d-1}(\mathbb{C}) \geqslant f(n)$. We say that $\mathbb{C}$ is an $f(n)$-**good covering** if $\varphi_R(\mathbb{C}) \leqslant f(n)$.*

## 3. Duality for a specific maximal code

A code $\mathbb{C} \subseteq \mathbb{F}_2^n$ is said to be ***maximal*** if it is not possible to adjoin any point of $\mathbb{F}_2^n$ to $\mathbb{C}$ without decreasing its minimum distance. Equivalently, a code $\mathbb{C}$ with minimum distance $d$ and covering radius $R$ is maximal if and only if $R \leqslant d-1$. Our first result is an easy theorem, which says that *any* maximal code is either a good packing or a good covering.

---

[*] The condition $d \leqslant 0.499n$ has been now improved to the more natural $d < n/2$ by Vu and Wu [9]. Vu and Wu [9] also show that a similar bound holds over any finite filed $\mathbb{F}_q$ provided $d < n(q-1)/q$.

**Theorem 1.** *Let $f(n)$ be an arbitrary function of $n$, and let $\mathbb{C} \subseteq \mathbb{F}_2^n$ be a maximal code. Then $\mathbb{C}$ is an $f(n)$-good packing or an $f(n)$-good covering (or both).*

*Proof.* By definition, $\mathbb{C}$ is not an $f(n)$-good packing if $\varphi_{d-1}(\mathbb{C}) < f(n)$. But this implies that $\varphi_R(\mathbb{C}) \leqslant \varphi_{d-1}(\mathbb{C}) < f(n)$, so $\mathbb{C}$ is an $f(n)$-good covering. ∎

For example, taking $f(n) = \theta(n)$, Theorem 1 implies that, up to a constant factor, any maximal code attains either the Jiang-Vardy bound (5) or the Delsarte-Piret bound (7).

## 4. Duality for almost all codes

We begin with three simple lemmas, which are needed to prove Theorems 5 and 6, our main results in this section. The following "supercode lemma" is well known.

**Lemma 2.** *Given a code $\mathbb{C}$, let $d(\mathbb{C})$ and $R(\mathbb{C})$ denote its minimum distance and covering radius, respectively. If $\mathbb{C}$ is a proper subcode of another code $\mathbb{C}'$, then $R(\mathbb{C}) \geqslant d(\mathbb{C}')$.*

*Proof.* Since $\mathbb{C} \subset \mathbb{C}'$, there exists an $x \in \mathbb{C}' \setminus \mathbb{C}$. For any $c \in \mathbb{C}$, we have $d(x,c) \geqslant d(\mathbb{C}')$. Hence $R(\mathbb{C}) \geqslant d(\mathbb{C}')$ by definition. ∎

**Lemma 3.** *Let $\mathcal{S}' \subseteq \mathscr{C}(n, M+1)$ be an arbitrary set of codes of length $n$ and size $M+1$, and let $\mathcal{S} = \{\mathbb{C} \in \mathscr{C}(n, M) : \mathbb{C} \subset \mathbb{C}' \text{ for some } \mathbb{C}' \in \mathcal{S}'\}$. Then the fraction of codes in $\mathcal{S}$ is greater or equal to the fraction of codes in $\mathcal{S}'$, namely*

$$\frac{|\mathcal{S}|}{|\mathscr{C}(n,M)|} \geqslant \frac{|\mathcal{S}'|}{|\mathscr{C}(n,M+1)|}$$

*Proof.* Define a bipartite graph $\mathcal{G}$ as follows. The left vertices, respectively the right vertices, of $\mathcal{G}$ are all the codes in $\mathscr{C}(n, M)$, respectively all the codes in $\mathscr{C}(n, M+1)$, with $\mathbb{C} \in \mathscr{C}(n, M)$ and $\mathbb{C}' \in \mathscr{C}(n, M+1)$ connected by an edge iff $\mathbb{C} \subset \mathbb{C}'$. Then $\mathcal{G}$ is bi-regular with left-degree $2^n - M$ and right-degree $M + 1$. Hence the number of edges in $\mathcal{G}$ is

$$|E(\mathcal{G})| = (M+1)|\mathscr{C}(n, M+1)| = (2^n - M)|\mathscr{C}(n, M)| \tag{8}$$

Now consider the subgraph $\mathcal{H}$ induced in $\mathcal{G}$ by the set $\mathcal{S}'$. Then the left vertices in $\mathcal{H}$ are precisely the codes in $\mathcal{S}$, and every such vertex has degree at most $2^n - M$. The degree of every right vertex in $\mathcal{H}$ is still $M + 1$. Thus, counting the number of edges in $\mathcal{H}$, we obtain

$$|E(\mathcal{H})| = (M+1)|\mathcal{S}'| \leqslant (2^n - M)|\mathcal{S}| \tag{9}$$

The lemma follows immediately from (8) and (9). Observe that the specific expressions for the left and right degrees of $\mathcal{G}$ are, in fact, irrelevant for the proof. ∎

**Lemma 4.** *Let $\mathcal{S}' \subseteq \mathscr{L}(n,k+1)$ be an arbitrary set of linear codes of length $n$ and dimension $k+1$, and let $\mathcal{S} = \{\mathbb{C} \in \mathscr{L}(n,k) : \mathbb{C} \subset \mathbb{C}' \text{ for some } \mathbb{C}' \in \mathcal{S}'\}$. Then the fraction of codes in $\mathcal{S}$ is greater or equal to the fraction of codes in $\mathcal{S}'$, namely*

$$\frac{|\mathcal{S}|}{|\mathscr{L}(n,k)|} \geqslant \frac{|\mathcal{S}'|}{|\mathscr{L}(n,k+1)|}$$

*Proof.* The argument is identical to the one given in the proof of Lemma 3, except that here we use the bipartite graph defined on $\mathscr{L}(n,k) \cup \mathscr{L}(n,k+1)$. ∎

The next theorem establishes the duality between the fraction of good coverings in $\mathscr{C}(n,M)$ and the fraction of good packings in $\mathscr{C}(n,M+1)$. In order to make its statement precise, we need to exclude the degenerate cases. Thus we henceforth assume that $n \leqslant M \leqslant 2^n - 1$.

**Theorem 5.** *Let $f(n)$ be an arbitrary function. Let $\alpha \in [0,1]$ denote the fraction of codes in $\mathscr{C}(n,M)$ that are $f(n)$-good coverings, and let $\beta \in [0,1]$ denote the fraction of codes in $\mathscr{C}(n,M+1)$ that are $f(n)$-good packings. Then $\alpha + \beta \leqslant 1$.*

*Proof.* Let $\mathcal{S}'$ be the set of all codes in $\mathscr{C}(n,M+1)$ that are $f(n)$-good packings. Thus $|\mathcal{S}'|/|\mathscr{C}(n,M+1)| = \beta$. Further, let $\mathcal{S} = \{\mathbb{C} \in \mathscr{C}(n,M) : \mathbb{C} \subset \mathbb{C}' \text{ for some } \mathbb{C}' \in \mathcal{S}'\}$ as in Lemma 3. We claim that none of the codes in $\mathcal{S}$ is an $f(n)$-good covering. Indeed, let $\mathbb{C} \in \mathcal{S}$, and let $\mathbb{C}' \in \mathcal{S}'$ be a code such that $\mathbb{C} \subset \mathbb{C}'$. Set $R = R(\mathbb{C})$ and $d = d(\mathbb{C}')$. Then

$$\varphi_R(\mathbb{C}) \geqslant \varphi_d(\mathbb{C}) \qquad \text{(by Lemma 2)} \tag{10}$$
$$> \varphi_{d-1}(\mathbb{C}') \qquad \text{(trivial from (1) if } M \geqslant n) \tag{11}$$
$$\geqslant f(n) \qquad (\mathbb{C}' \text{ is an } f(n)\text{-good packing}) \tag{12}$$

Thus $\mathbb{C}$ is not an $f(n)$-good covering, as claimed. Hence $1 - \alpha \geqslant |\mathcal{S}|/|\mathscr{C}(n,M)|$. The theorem now follows immediately from Lemma 3. ∎

For linear codes, exactly the same argument works, except that we need a factor of 2 in (11), since $|\mathbb{C}'| = 2|\mathbb{C}|$ for any $\mathbb{C} \in \mathscr{L}(n,k)$ and $\mathbb{C}' \in \mathscr{L}(n,k+1)$. For the functions $f(n)$ of the kind one usually considers, such constant factors are not particularly significant.

**Theorem 6.** *Let $f(n)$ be an arbitrary function. Let $\alpha \in [0,1]$ denote the fraction of codes in $\mathscr{L}(n,k)$ that are $f(n)$-good coverings, and let $\beta \in [0,1]$ denote the fraction of codes in $\mathscr{L}(n,k+1)$ that are $2f(n)$-good packings. Then $\alpha + \beta \leqslant 1$.*

*Proof.* Follows from Lemmas 2 and 4 in the same way as Theorem 5 follows from Lemmas 2 and 3. Explicitly, (10)–(12) becomes $\varphi_R(\mathbb{C}) \geqslant \varphi_d(\mathbb{C}) > \tfrac{1}{2}\varphi_{d-1}(\mathbb{C}') \geqslant f(n)$. ∎

## 5. Discussion

Clearly, Theorems 5 and 6 imply the statements ✻ and ✺, respectively, made in §1. If $\alpha$ tends to one as $n \to \infty$, then $\beta$ tends to zero, and vice versa if $\beta \to 1$ then $\alpha \to 0$.

It is well known [8] that almost all linear* codes achieve the Gilbert-Varshamov bound (4). Hence an intriguing question is what fraction of codes in $\mathscr{L}(n,k)$ achieve the stronger bound (5) of Jiang and Vardy [7]. Combining Theorem 6 with the results of Blinovskii [2] on random *covering* codes establishes the following theorem.

**Theorem 7.** *Let $n$ and $k = \lambda n$ be positive integers, with $0 < \lambda < 1$. For any real $\varepsilon > 0$, let $\beta_\varepsilon(n,k)$ denote the fraction of codes in $\mathscr{L}(n,k)$ whose minimum distance $d$ is such that $\varphi_{d-1}(\mathbb{C}) \geqslant n^{1+\varepsilon}$. Then $\beta_\varepsilon(n,k)$ tends to zero as $n \to \infty$, for all $\varepsilon$ and $\lambda$.*

We omit the proof of Theorem 7, since Dumer [6] recently proved a stronger result. Specifically, Dumer [6] shows that the fraction of linear codes that are $f(n)$-good packings tends to zero as $n \to \infty$ for *any* function $f(n)$ such that $\lim_{n\to\infty} f(n) = \infty$. This implies that as $n \to \infty$, almost all linear codes satisfy $\varphi_{d-1}(\mathbb{C}) = \theta(1)$.

**Acknowledgment.** We are grateful to Alexander Barg and Ilya Dumer for helpful discussions. We are especially indebted to Ilya Dumer for sending us his proof in [6].

---

*It is also known [1] that almost all nonlinear codes do *not* achieve the Gilbert-Varshamov bound.

# Duality between Packings and Coverings of the Hamming Space

**Gérard Cohen**

Département Informatique
Ecole Nationale Supérieure des Télécommunications
46 rue Barrault, 75634 Paris, FRANCE
*cohen@enst.fr*

**Alexander Vardy**

Department of Electrical and Computer Engineering
Department of Computer Science and Engineering
Department of Mathematics
University of California San Diego
9500 Gilman Drive, La Jolla, CA 92093, U.S.A.
*vardy@kilimanjaro.ucsd.edu*

July 4, 2005

*Dedicated to the memory of Jack van Lint*

**Abstract**

We investigate the packing and covering densities of linear and nonlinear binary codes, and establish a number of duality relationships between the packing and covering problems. Specifically, we prove that if almost all codes (in the class of linear or nonlinear codes) are good packings, then only a vanishing fraction of codes are good coverings, and vice versa: if almost all codes are good coverings, then at most a vanishing fraction of codes are good packings. We also show that any *specific* maximal binary code is either a good packing or a good covering, in a certain well-defined sense.

Supported in part by the David and Lucile Packard Fellowship and by the National Science Foundation.

# 1. Introduction

Let $\mathbb{F}_2^n$ be the vector space of all the binary $n$-tuples, endowed with the Hamming metric. Specifically, the **Hamming distance** $d(x,y)$ between $x, y \in \mathbb{F}_2^n$ is defined as the number of positions where $x$ and $y$ differ. A **binary code** of length $n$ is a subset of $\mathbb{F}_2^n$, while a **binary linear code** of length $n$ and dimension $k$ is a $k$-dimensional subspace of $\mathbb{F}_2^n$. Since we are concerned *only* with binary codes in this paper, we henceforth omit the "binary" quantifier throughout. The **minimum distance** $d$ of a code $\mathbb{C} \subseteq \mathbb{F}_2^n$ is defined as the minimum Hamming distance between distinct elements of $\mathbb{C}$. The **covering radius** of $\mathbb{C}$ is the smallest integer $R$ such that for all $x \in \mathbb{F}_2^n$, there is a $y \in \mathbb{C}$ with $d(x,y) \leqslant R$. For all other notation from coding theory, we refer the reader to the book of van Lint [8]. Van Lint [8, p.34] calls the covering radius the "counterpart of minimum distance." Indeed, the trade-off between the parameters $|\mathbb{C}|$, $n$, $d$, and $R$ is one of the fundamental problems in coding theory.

Let $\mathscr{C}(n, M)$ denote the set of all codes $\mathbb{C} \subseteq \mathbb{F}_2^n$ with $|\mathbb{C}| = M$. Thus $|\mathscr{C}(n, M)| = \binom{2^n}{M}$. Similarly, let $\mathscr{L}(n,k)$ denote the set of all linear codes of length $n$ and dimension $k$. Thus the cardinality of $\mathscr{L}(n,k)$ is given by $|\mathscr{L}(n,k)| = \prod_{i=0}^{k-1} (2^n - 2^i)/(2^k - 2^i)$. We will be interested in questions of the following kind. Given a property **P** which determines the packing or covering density of a code, what fraction of codes in $\mathscr{C}(n,M)$ and/or $\mathscr{L}(n,k)$ have this property? Moreover, how does this fraction behave as $n \to \infty$? Our main results are curious duality relationships between such packing and covering problems. In particular, we show that:

- ✧ *Any maximal code is good. That is, any specific maximal code $\mathbb{C} \subseteq \mathbb{F}_2^n$ is either a good packing or a good covering, in a certain well-defined sense.*

- ✷ *If almost all codes in $\mathscr{C}(n,M)$ are good coverings, then almost all codes in $\mathscr{C}(n,M+1)$ are bad packings. Vice versa, if almost all codes in $\mathscr{C}(n,M+1)$ are good packings, then almost all codes in $\mathscr{C}(n,M)$ are bad coverings.*

- ✷ *The same is true for linear codes. That is, ✷ holds with $\mathscr{C}(n,M)$ and $\mathscr{C}(n,M+1)$ replaced by $\mathscr{L}(n,k)$ and $\mathscr{L}(n,k+1)$, respectively.*

The definition of what we mean by "good packing" and "good covering" is given in the next section. Precise statements of ✧ and ✷, ✷ may be found in §3 and §4, respectively.

# 2. Definitions

The **covering density** of a code $\mathbb{C} \subseteq \mathbb{F}_2^n$ is defined in [3] as the sum of the volumes of spheres of covering radius $R$ about the codewords of $\mathbb{C}$ divided by the volume of the space:

$$\mu(\mathbb{C}) \stackrel{\text{def}}{=} \frac{\sum_{c \in \mathbb{C}} |B_R(c)|}{|\mathbb{F}_2^n|} = \frac{|\mathbb{C}| V(n,R)}{2^n}$$

where $B_r(x) = \{y \in \mathbb{F}_2^n : d(x,y) \leqslant r\}$ is the sphere (ball) of radius $r$ centered at $x \in \mathbb{F}_2^n$ and $V(n,r) = \sum_{i=0}^{r} \binom{n}{i}$ is the volume (cardinality) of $B_r(x)$. We find it extremely convenient to extend this definition of density to arbitrary radii as follows.

**Definition 1.** *Given a code $\mathbb{C} \subseteq \mathbb{F}_2^n$ and a nonnegative integer $r \leqslant n$, the r-**density of** $\mathbb{C}$ is defined as*

$$\varphi_r(\mathbb{C}) \stackrel{\text{def}}{=} \frac{\sum_{c \in \mathbb{C}} |B_r(c)|}{|\mathbb{F}_2^n|} = \frac{|\mathbb{C}| V(n,r)}{2^n} \quad (1)$$

Many well-known bounds on the packing and covering density of codes can be concisely stated in terms of the $r$-density. For example, if $R$, $d$, and $t = \lfloor (d-1)/2 \rfloor$ denote the covering radius, the minimum distance, and the packing radius, respectively, then

*Sphere-packing bound:* $\varphi_t(\mathbb{C}) \leqslant 1$ for all $\mathbb{C} \subseteq \mathbb{F}_2^n$ \quad (2)

*Sphere-covering bound:* $\varphi_R(\mathbb{C}) \geqslant 1$ for all $\mathbb{C} \subseteq \mathbb{F}_2^n$ \quad (3)

The classical Gilbert-Varshamov bound [8] asserts that for all $n$ and $d \leqslant n$, there exist codes in $\mathscr{C}(n, M)$ whose minimum distance $d$ satisfies $M \geqslant 2^n/V(n, d-1)$. Equivalently

*Gilbert-Varshamov bound:* $\forall n, \forall d \leqslant n$, there exist $\mathbb{C} \subseteq \mathbb{F}_2^n$, such that $\varphi_{d-1}(\mathbb{C}) \geqslant 1$ \quad (4)

Recently, this bound was improved upon by Jiang and Vardy [7] who proved that for all sufficiently large $n$ and all[*] $d \leqslant 0.499n$, there exist codes $\mathbb{C} \subset \mathbb{F}_2^n$ with minimum distance $d$ such that $|\mathbb{C}| \geqslant cn\, 2^n/V(n, d-1)$, where $c$ is an absolute constant. Equivalently

$$\exists c > 0, \forall n \geqslant n_0, \forall d \leqslant 0.499n, \text{ there exist } \mathbb{C} \subseteq \mathbb{F}_2^n, \text{ such that } \varphi_{d-1}(\mathbb{C}) \geqslant cn \quad (5)$$

The best known existence bounds for covering codes can be also expressed in terms of the $r$-density, except that one should set $r = R$ rather than $r = d-1$. Thus

$\forall n, \forall R < n/2$, there exist linear $\mathbb{C} \subseteq \mathbb{F}_2^n$, such that $\varphi_R(\mathbb{C}) \leqslant n^2$ \quad (6)

$\forall n, \forall R < n/2$, there exist $\mathbb{C} \subseteq \mathbb{F}_2^n$, such that $\varphi_R(\mathbb{C}) \leqslant (\ln 2)n$ \quad (7) where the first result is due to Cohen [4] while the second is due to Delsarte and Piret [5]. Motivated by (4)–(7), we introduce the following definition.

**Definition 2.** *Let $f(n)$ be a given function, and let $\mathbb{C} \subseteq \mathbb{F}_2^n$ be a code with minimum distance $d$ and covering radius $R$. We say that $\mathbb{C}$ is an $f(n)$-**good packing** if $\varphi_{d-1}(\mathbb{C}) \geqslant f(n)$. We say that $\mathbb{C}$ is an $f(n)$-**good covering** if $\varphi_R(\mathbb{C}) \leqslant f(n)$.*

## 3. Duality for a specific maximal code

A code $\mathbb{C} \subseteq \mathbb{F}_2^n$ is said to be **maximal** if it is not possible to adjoin any point of $\mathbb{F}_2^n$ to $\mathbb{C}$ without decreasing its minimum distance. Equivalently, a code $\mathbb{C}$ with minimum distance $d$ and covering radius $R$ is maximal if and only if $R \leqslant d - 1$. Our first result is an easy theorem, which says that *any* maximal code is either a good packing or a good covering.

---

[*] The condition $d \leqslant 0.499n$ has been now improved to the more natural $d < n/2$ by Vu and Wu [9]. Vu and Wu [9] also show that a similar bound holds over any finite filed $\mathbb{F}_q$ provided $d < n(q-1)/q$.

**Theorem 1.** *Let $f(n)$ be an arbitrary function of $n$, and let $\mathbb{C} \subseteq \mathbb{F}_2^n$ be a maximal code. Then $\mathbb{C}$ is an $f(n)$-good packing or an $f(n)$-good covering (or both).*

*Proof.* By definition, $\mathbb{C}$ is not an $f(n)$-good packing if $\varphi_{d-1}(\mathbb{C}) < f(n)$. But this implies that $\varphi_R(\mathbb{C}) \leqslant \varphi_{d-1}(\mathbb{C}) < f(n)$, so $\mathbb{C}$ is an $f(n)$-good covering. ∎

For example, taking $f(n) = \theta(n)$, Theorem 1 implies that, up to a constant factor, any maximal code attains either the Jiang-Vardy bound (5) or the Delsarte-Piret bound (7).

## 4. Duality for almost all codes

We begin with three simple lemmas, which are needed to prove Theorems 5 and 6, our main results in this section. The following "supercode lemma" is well known.

**Lemma 2.** *Given a code $\mathbb{C}$, let $d(\mathbb{C})$ and $R(\mathbb{C})$ denote its minimum distance and covering radius, respectively. If $\mathbb{C}$ is a proper subcode of another code $\mathbb{C}'$, then $R(\mathbb{C}) \geqslant d(\mathbb{C}')$.*

*Proof.* Since $\mathbb{C} \subset \mathbb{C}'$, there exists an $x \in \mathbb{C}' \setminus \mathbb{C}$. For any $c \in \mathbb{C}$, we have $d(x,c) \geqslant d(\mathbb{C}')$. Hence $R(\mathbb{C}) \geqslant d(\mathbb{C}')$ by definition. ∎

**Lemma 3.** *Let $\mathcal{S}' \subseteq \mathscr{C}(n, M+1)$ be an arbitrary set of codes of length $n$ and size $M + 1$, and let $\mathcal{S} = \{\mathbb{C} \in \mathscr{C}(n, M) : \mathbb{C} \subset \mathbb{C}' \text{ for some } \mathbb{C}' \in \mathcal{S}'\}$. Then the fraction of codes in $\mathcal{S}$ is greater or equal to the fraction of codes in $\mathcal{S}'$, namely*

$$\frac{|\mathcal{S}|}{|\mathscr{C}(n, M)|} \geqslant \frac{|\mathcal{S}'|}{|\mathscr{C}(n, M+1)|}$$

*Proof.* Define a bipartite graph $\mathcal{G}$ as follows. The left vertices, respectively the right vertices, of $\mathcal{G}$ are all the codes in $\mathscr{C}(n, M)$, respectively all the codes in $\mathscr{C}(n, M+1)$, with $\mathbb{C} \in \mathscr{C}(n, M)$ and $\mathbb{C}' \in \mathscr{C}(n, M+1)$ connected by an edge iff $\mathbb{C} \subset \mathbb{C}'$. Then $\mathcal{G}$ is bi-regular with left-degree $2^n - M$ and right-degree $M + 1$. Hence the number of edges in $\mathcal{G}$ is

$$|E(\mathcal{G})| = (M+1)|\mathscr{C}(n, M+1)| = (2^n - M)|\mathscr{C}(n, M)| \tag{8}$$

Now consider the subgraph $\mathcal{H}$ induced in $\mathcal{G}$ by the set $\mathcal{S}'$. Then the left vertices in $\mathcal{H}$ are precisely the codes in $\mathcal{S}$, and every such vertex has degree at most $2^n - M$. The degree of every right vertex in $\mathcal{H}$ is still $M + 1$. Thus, counting the number of edges in $\mathcal{H}$, we obtain

$$|E(\mathcal{H})| = (M+1)|\mathcal{S}'| \leqslant (2^n - M)|\mathcal{S}| \tag{9}$$

The lemma follows immediately from (8) and (9). Observe that the specific expressions for the left and right degrees of $\mathcal{G}$ are, in fact, irrelevant for the proof. ∎

**Lemma 4.** *Let $\mathcal{S}' \subseteq \mathscr{L}(n, k+1)$ be an arbitrary set of linear codes of length $n$ and dimension $k + 1$, and let $\mathcal{S} = \{\mathbb{C} \in \mathscr{L}(n, k) : \mathbb{C} \subset \mathbb{C}' \text{ for some } \mathbb{C}' \in \mathcal{S}'\}$. Then the fraction of codes in $\mathcal{S}$ is greater or equal to the fraction of codes in $\mathcal{S}'$, namely*

$$\frac{|\mathcal{S}|}{|\mathscr{L}(n, k)|} \geqslant \frac{|\mathcal{S}'|}{|\mathscr{L}(n, k+1)|}$$

*Proof.* The argument is identical to the one given in the proof of Lemma 3, except that here we use the bipartite graph defined on $\mathscr{L}(n, k) \cup \mathscr{L}(n, k+1)$. ∎

The next theorem establishes the duality between the fraction of good coverings in $\mathscr{C}(n,M)$ and the fraction of good packings in $\mathscr{C}(n,M+1)$. In order to make its statement precise, we need to exclude the degenerate cases. Thus we henceforth assume that $n \leqslant M \leqslant 2^n - 1$.

**Theorem 5.** *Let $f(n)$ be an arbitrary function. Let $\alpha \in [0,1]$ denote the fraction of codes in $\mathscr{C}(n,M)$ that are $f(n)$-good coverings, and let $\beta \in [0,1]$ denote the fraction of codes in $\mathscr{C}(n,M+1)$ that are $f(n)$-good packings. Then $\alpha + \beta \leqslant 1$.*

*Proof.* Let $\mathcal{S}'$ be the set of all codes in $\mathscr{C}(n,M+1)$ that are $f(n)$-good packings. Thus $|\mathcal{S}'|/|\mathscr{C}(n,M+1)| = \beta$. Further, let $\mathcal{S} = \{\mathbb{C} \in \mathscr{C}(n,M) : \mathbb{C} \subset \mathbb{C}' \text{ for some } \mathbb{C}' \in \mathcal{S}'\}$ as in Lemma 3. We claim that none of the codes in $\mathcal{S}$ is an $f(n)$-good covering. Indeed, let $\mathbb{C} \in \mathcal{S}$, and let $\mathbb{C}' \in \mathcal{S}'$ be a code such that $\mathbb{C} \subset \mathbb{C}'$. Set $R = R(\mathbb{C})$ and $d = d(\mathbb{C}')$. Then

$$\varphi_R(\mathbb{C}) \geqslant \varphi_d(\mathbb{C}) \qquad \text{(by Lemma 2)} \qquad (10)$$
$$> \varphi_{d-1}(\mathbb{C}') \qquad \text{(trivial from (1) if } M \geqslant n\text{)} \qquad (11)$$
$$\geqslant f(n) \qquad \text{(}\mathbb{C}' \text{ is an } f(n)\text{-good packing)} \qquad (12)$$

Thus $\mathbb{C}$ is not an $f(n)$-good covering, as claimed. Hence $1 - \alpha \geqslant |\mathcal{S}|/|\mathscr{C}(n,M)|$. The theorem now follows immediately from Lemma 3. ∎

For linear codes, exactly the same argument works, except that we need a factor of 2 in (11), since $|\mathbb{C}'| = 2|\mathbb{C}|$ for any $\mathbb{C} \in \mathscr{L}(n,k)$ and $\mathbb{C}' \in \mathscr{L}(n,k+1)$. For the functions $f(n)$ of the kind one usually considers, such constant factors are not particularly significant.

**Theorem 6.** *Let $f(n)$ be an arbitrary function. Let $\alpha \in [0,1]$ denote the fraction of codes in $\mathscr{L}(n,k)$ that are $f(n)$-good coverings, and let $\beta \in [0,1]$ denote the fraction of codes in $\mathscr{L}(n,k+1)$ that are $2f(n)$-good packings. Then $\alpha + \beta \leqslant 1$.*

*Proof.* Follows from Lemmas 2 and 4 in the same way as Theorem 5 follows from Lemmas 2 and 3. Explicitly, (10)–(12) becomes $\varphi_R(\mathbb{C}) \geqslant \varphi_d(\mathbb{C}) > \frac{1}{2}\varphi_{d-1}(\mathbb{C}') \geqslant f(n)$. ∎

## 5. Discussion

Clearly, Theorems 5 and 6 imply the statements ✸ and ✹, respectively, made in §1. If $\alpha$ tends to one as $n \to \infty$, then $\beta$ tends to zero, and vice versa if $\beta \to 1$ then $\alpha \to 0$.

It is well known [8] that almost all linear* codes achieve the Gilbert-Varshamov bound (4). Hence an intriguing question is what fraction of codes in $\mathscr{L}(n,k)$ achieve the stronger bound (5) of Jiang and Vardy [7]. Combining Theorem 6 with the results of Blinovskii [2] on random *covering* codes establishes the following theorem.

**Theorem 7.** *Let $n$ and $k = \lambda n$ be positive integers, with $0 < \lambda < 1$. For any real $\varepsilon > 0$, let $\beta_\varepsilon(n,k)$ denote the fraction of codes in $\mathscr{L}(n,k)$ whose minimum distance $d$ is such that $\varphi_{d-1}(\mathbb{C}) \geqslant n^{1+\varepsilon}$. Then $\beta_\varepsilon(n,k)$ tends to zero as $n \to \infty$, for all $\varepsilon$ and $\lambda$.*

---

*It is also known [1] that almost all nonlinear codes do *not* achieve the Gilbert-Varshamov bound.

We omit the proof of Theorem 7, since Dumer [6] recently proved a stronger result. Specifically, Dumer [6] shows that the fraction of linear codes that are $f(n)$-good packings tends to zero as $n \to \infty$ for *any* function $f(n)$ such that $\lim_{n\to\infty} f(n) = \infty$. This implies that as $n \to \infty$, almost all linear codes satisfy $\varphi_{d-1}(\mathbb{C}) = \theta(1)$.

**Acknowledgment.** We are grateful to Alexander Barg and Ilya Dumer for helpful discussions. We are especially indebted to Ilya Dumer for sending us his proof in [6].